\documentclass[11pt,twoside]{article}
\usepackage{asp2010}

\resetcounters

\begin{document}

\title{Inverting the White Dwarf Luminosity Function: the Star Formation History of the Solar Neighbourhood}
\author{Nicholas~Rowell
\affil{Space Technology Centre, School of Computing, University of Dundee, Dundee, UK}}

\begin{abstract}
I present an algorithm for inverting the luminosity function for white dwarfs to obtain a maximum likelihood estimate
of the star formation rate of the host stellar population. The algorithm is of the general class of Expectation
Maximization, and involves iteratively improving an initial guess of the star formation rate.
Tests show that the inversion results are quite sensitive to the assumed metallicity and initial mass function, but relatively
insensitive to the initial-final mass relation and ratio of H/He atmosphere white dwarfs.
Application to two independent determinations of the Solar neighbourhood white dwarf luminosity function
gives similar results: the star formation rate is characterised by an early burst, and more recent peak at 2-3 Gyr
in the past.
\end{abstract}

\section{Introduction}
The white dwarf luminosity function (WDLF) is a useful tool for determining the age of a population of stars.
The magnitude at which the function terminates is highly time-dependent,
and by fitting the faint end with theoretical WDLF models
of different ages one can obtain a statistical estimate of the age of the population without having to
determine the total age of any individual white dwarf, which is considerably more difficult.
This technique has been applied both to single burst populations such as open clusters \citep{Bedin2010, Garcia-Berro2010}
and continuous populations such as the Galactic disk \citep{Oswalt1996,Knox1999}.

The standard equation for modelling the WDLF for a given star formation history is \citep[e.g.][]{Iben1989,Fontaine2001}
\begin{equation}
\Phi(M_{\mathrm{bol}}) = \int\limits^{M_u}\limits_{M_l} \frac{dt_{\mathrm{cool}}}{dM_{\mathrm{bol}}} 
\, \psi(T_0 - t_{\mathrm{cool}} - t_{\mathrm{MS}}) \, \phi(M) \, dM      
\label{eq:theoryLF}                             
\end{equation}
where $\Phi(M_{\mathrm{bol}})$ is the number density of WDs at magnitude $M_{\mathrm{bol}}$.
The derivative inside the integral is the characteristic cooling time for WDs, $\psi(t)$ is the star formation rate (SFR) at time $t$
and $\phi$ is the initial mass function (IMF). The integral also depends on the lifetimes of main sequence progenitors as a function
of mass and metallicity $t_{\mathrm{MS}}$,
the WD cooling times as a function of mass and luminosity $t_{\mathrm{cool}}$, the initial-final mass relation $m(M)$ and the total
time since the onset of star formation $T_0$. The integral is over all main sequence masses that have had time to produce WDs at the present
day, with the magnitude-dependent lower limit corresponding to the solution of
\[
T_0 - t_{\mathrm{cool}}(M_{\mathrm{bol}},m(M_l)) - t_{\mathrm{MS}}(M_l,Z) = 0
\]
and the upper limit $M_u \approx 7$ M$_{\odot}$.

From various studies of this equation \citep{Iben1989,Noh1990} it is known that the faint end of the WDLF is mostly
insensitive to the SFR, and is determined mainly by the total age of the population $T_0$. WDs at these magnitudes
are uniformly old, and are the remains of high mass main sequence stars that formed right at the onset of star formation.
It is for this reason that the faint end provides the most constraint on the total age.
The picture is considerably more complicated at brighter magnitudes, because the WDs are a mixture of ages: both young, high mass WDs that are
produced by recently-formed MS progenitors, and old, low mass WDs that are produced by low mass MS stars that formed at early times. 
It was found by \citet{Noh1990} that time variations in the SFR may leave imprints in the WDLF at these magnitudes,
and by forward modelling methods they interpreted a marginal feature in the WDLF at $M_{\mathrm{bol}} \approx 10$
as evidence for a burst of star formation $0.3$ Gyr ago.
According to \citet{Noh1990} and equation \ref{eq:theoryLF}, the shape of the WDLF at intermediate magnitudes is also strongly 
affected by the cooling rates of WDs, and it is possible that features in the WDLF may be interpreted as evidence of additional 
WD cooling mechanisms \citep[see e.g.][ and the contribution of Miguel \& Bertolami to these proceedings]{Isern2008}.

This paper presents results of ongoing work on a strategy to invert the WDLF to obtain a direct estimate of the
time varying SFR. This work is driven by two related questions: given current WD cooling models, what constraint 
can features in the WDLF (at all magnitudes)
place on the time varying SFR? And as a corrolary to this: can features in the WDLF be
explained exclusively by time variations in the SFR, or are additional cooling mechanisms required?
\section{White Dwarf Luminosity Function Inversion Algorithm}
To a first approximation, the two parameters that determine the total age of a WD are the present day bolometric magnitude, and the mass.
These can be used to determine both the total WD cooling time and the time spent on the main sequence.
The approach to inverting the WDLF presented here is based on the observation that if the distribution of WD mass was known at all magnitudes,
then the WDLF could be immediately transformed to the SFR. As this quantity is generally not known observationally,
this direct approach can't be used. Instead, we use the inversion technique known as Expectation Maximization \citep{Dempster1977,Do2008},
which is used to obtain maximum likelihood estimates of the solution to inverse problems in the presence of missing data.

This approach involves iteratively refining an initial guess of the SFR. The general procedure for each iteration is as follows.
The starting point is an initial guess of the star formation rate $\psi_0$,
\begin{equation}
\psi_0  \equiv \psi_0(t),
\end{equation}
where in the present work $\psi_0$ is flat, i.e. a constant star formation rate.
This is combined with the initial mass function $\phi$ to get the joint mass and formation time distribution of main sequence progenitors
$P_{\mathrm{MS}}$, where
\begin{equation}
P_{\mathrm{MS}}(M_{\mathrm{MS}}, t) = \phi(M_{\mathrm{MS}}) \psi_0(t)
\end{equation}
Using standard rules of probability density functions, we can transform this to the joint mass and bolometric magnitude 
distribution of white dwarfs $P_{\mathrm{WD}}$ at the present day:
\begin{equation}
\label{eq:wdDist}
P_{\mathrm{WD}}(M_{\mathrm{WD}}, M_{\mathrm{bol}}) = P_{\mathrm{MS}}(M_{\mathrm{MS}}, t). \left| \frac{\partial(M_{\mathrm{MS}},t)}{\partial(M_{\mathrm{WD}},M_{\mathrm{bol}})} \right|
\end{equation}
This function can be seperated into a product of the marginal luminosity distribution
and the mass distribution conditioned on luminosity,
\begin{equation}
P_{\mathrm{WD}}(M_{\mathrm{WD}}, M_{\mathrm{bol}}) = \Phi_{\mathrm{sim}}(M_{\mathrm{bol}}) P_{\mathrm{WD}}(M_{\mathrm{WD}} | M_{\mathrm{bol}})
\end{equation}
The quantity $\Phi_{\mathrm{sim}}$ is just the WDLF for the initial guess SFR model, up to a normalisation factor.
The next crucial step is to replace this with the observed WDLF $\Phi_{\mathrm{obs}}$ to get the updated WD distribution $P_{\mathrm{WD}}^{\prime}$:
\begin{equation}
P_{\mathrm{WD}}^{\prime}(M_{\mathrm{WD}}, M_{\mathrm{bol}}) = \Phi_{\mathrm{obs}}(M_{\mathrm{bol}}) P_{\mathrm{WD}}(M_{\mathrm{WD}} | M_{\mathrm{bol}})
\end{equation}
This updated WD distribution has the same marginal luminosity distribution as the observed WDLF, and the magnitude-dependent mass 
distribution derived from the initial guess star formation rate model.
We can now invert this distribution to obtain the updated distribution for main sequence stars $P_{\mathrm{MS}}^{\prime}$
again using standard transformation rules:
\begin{equation}
P_{\mathrm{MS}}^{\prime}(M_{\mathrm{MS}}, t) = P_{\mathrm{WD}}^{\prime}(M_{\mathrm{WD}}, M_{\mathrm{bol}}) \left| \frac{\partial(M_{\mathrm{WD}},M_{\mathrm{bol}})}{\partial(M_{\mathrm{MS}},t)} \right|
\end{equation}
The final step is to marginalise $P_{\mathrm{MS}}^{\prime}$ over the main sequence mass, to obtain the updated 
star formation rate model $\psi_1$:
\begin{equation}
\psi_1(t)  = \frac{1}{1-A(t)} \int_{M_{\mathrm{MS}}^{\mathrm{lifetime}}(t)}^{M_{\mathrm{MS}}^{\mathrm{max}}} \! P_{\mathrm{MS}}^{\prime}(M_{\mathrm{MS}}, t) \, \mathrm{d} M_{\mathrm{MS}}
\end{equation}
The integral is over all main sequence stars that produce WDs at the present day. In the present work, the upper limit 
$M_{\mathrm{MS}}^{\mathrm{max}}=7$M$_{\odot}$, and the variable lower limit $M_{\mathrm{MS}}^{\mathrm{lifetime}}(t)$ correponds to the 
mass of the main sequence star with lifetime $t$.
The factor $A$ corrects for low mass MS stars that don't form WDs at the present day, and is calculated
\begin{equation}
A(t) = \int_{M_{\mathrm{MS}}^{\mathrm{min}}}^{M_{\mathrm{MS}}^{\mathrm{lifetime}}(t)} \! \phi(M_{\mathrm{MS}}) \, \mathrm{d} M_{\mathrm{MS}}
\end{equation}
where the lower mass limit in this case is set to $M_{\mathrm{MS}}^{\mathrm{min}}=0.6$ M$_{\odot}$. The star formation rate recovered by this
algorithm therefore only represents stars more massive than $0.6$ M$_{\odot}$.
It is also non-parametric, in the sense that
it does not enforce any particular functional form on the recovered SFR.
\subsection{White Dwarf Atmosphere Types}
Along with the mass and present luminosity, the H/He atmosphere type is a third parameter affecting the total age of a WD.
This has a significant effect at larger cooling ages ($\gtrsim6$ Gyr depending on the choice of models), with H atmosphere 
WDs being brighter at a given cooling age.
The two atmosphere types can be included in the algorithm in a relatively straightforward manner. We calculate
$P_{\mathrm{WD}}$ in equation \ref{eq:wdDist} separately for each atmosphere type, then take a linear combination to obtain
the total $P_{\mathrm{WD}}$ for the mixed atmosphere population, where
\begin{equation}
P_{\mathrm{WD}} = \alpha P_{\mathrm{WD}}^{H} + (1-\alpha) P_{\mathrm{WD}}^{He}
\end{equation}
The factor $\alpha$ fixes the relative abundance of H and He WDs at birth, though their ratio changes with luminosity
due to the two types cooling at different rates. A value of $\alpha=0.5$ is used in the present work.
\section{Validation with Synthetic Data}
\begin{figure}[h!]
\plottwo{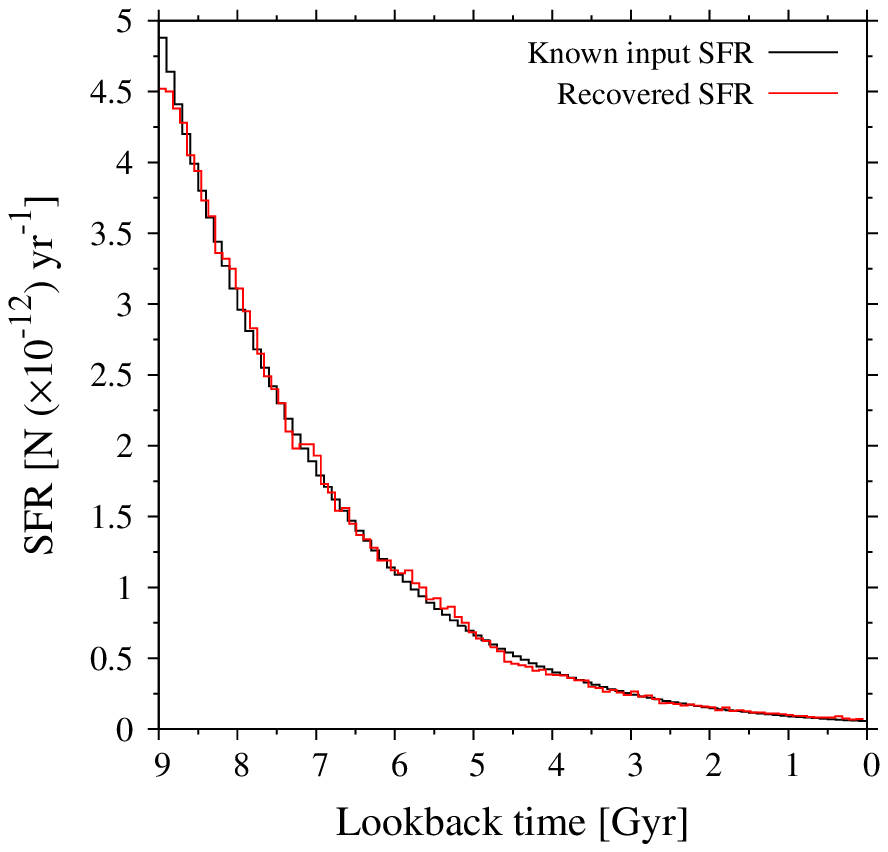}{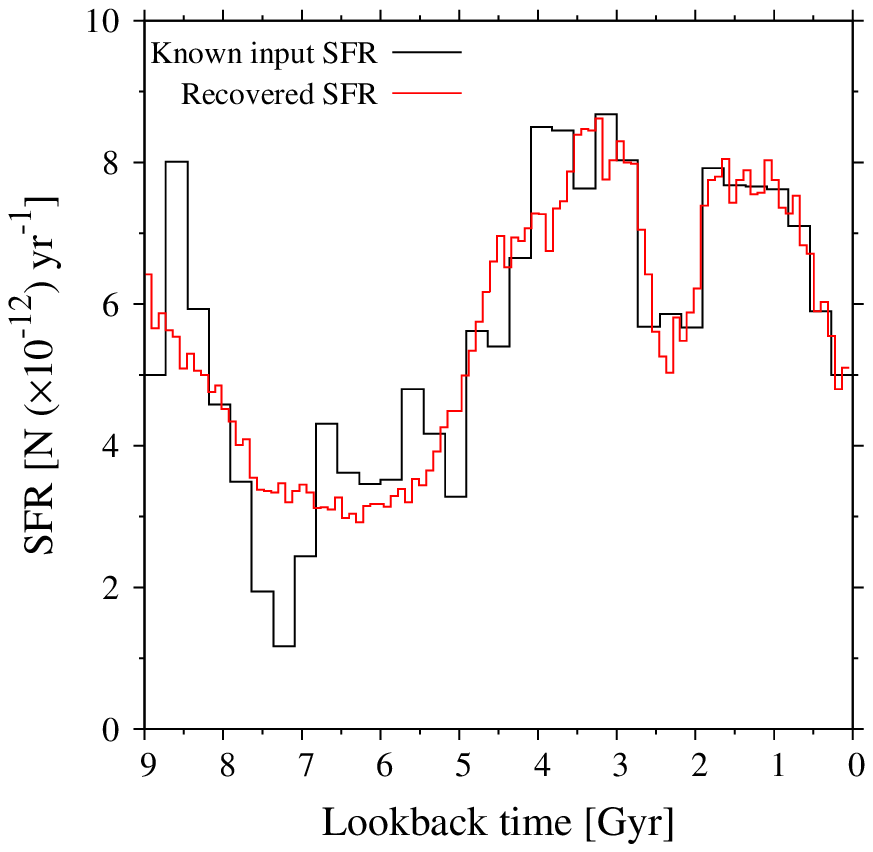}
\caption{Testing convergence of inversion algorithm using synthetic WDLF data. The black lines show the underlying SFR model used to 
generate the artificial WDLF in each case; the red lines show the recovered SFR model on the final iteration of the algorithm.
In these tests, the algorithm converged in 27 and 32 steps.}
\label{fig:verification}
\end{figure}

In order to test the accuracy of the recovered SFR, we have generated a set of synthetic WDLFs using a range
of different known input SFR models.
This allows us to check the performance of the algorithm in tightly controlled 
noise conditions, and the sensitivity to uncertainties in the various modelling inputs.
In this work we use the WD cooling sequences described in \citet{Tremblay2011}
and \citet{Bergeron2011} and references therein (see also \texttt{www.astro.umontreal.ca/$\sim$bergeron/CoolingModels}).
Figure \ref{fig:verification} shows the results from two tests in noise-free conditions, designed as a proof of concept
to verify that the algorithm works in principle on realistic models of the SFR. In each case, the synthetic WDLF (not shown) 
has a magnitude binning of $\Delta M_{\mathrm{bol}}=0.5$, chosen to match the observed WDLF resolution in recent studies.
The algorithm performs well on smoothly varying SFRs like the exponential decay model (left). 
The overall form of the fractal SFR model on the right is recovered by
the inversion, although at older times high frequency components in the underlying SFR are lost and the algorithm only
measures a moving average. This is a fundamental limit of the algorithm arising from the finite
magnitude resolution in the WDLF.

Inverse problems are notoriously sensitive to noise. We have carried out a comprehensive campaign of tests to assess the
effect of observational errors and modelling uncertainties on the algorithm performance. To summarise, observational
errors on the scale of recent WDLF measurements in the Solar neighbourhood do not catastrophically degrade the 
performance of the algorithm, and we are able to correctly estimate the true error on the inverted SFR.
In terms of modelling parameters, the algorithm is most sensitive to uncertainties in the IMF and progenitor metallicity,
and less sensitive to the initial-final mass relation and H/He atmosphere ratio. The full results will be published
separately.
\section{The Solar Neighbourhood}
\begin{figure}[h!]
\plottwo{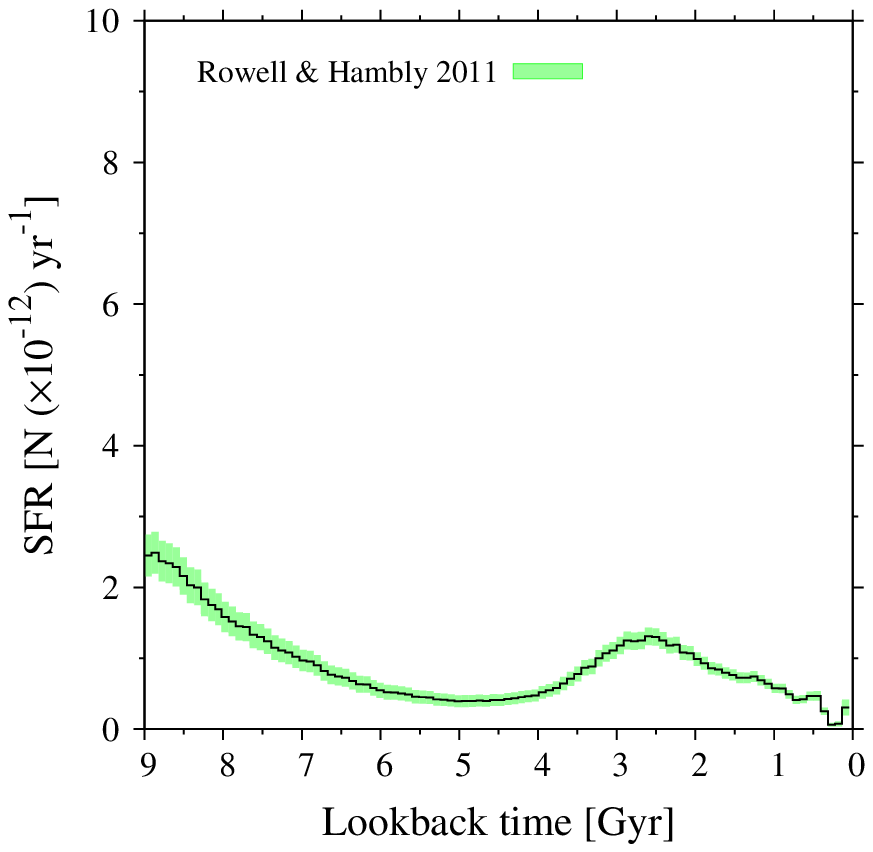}{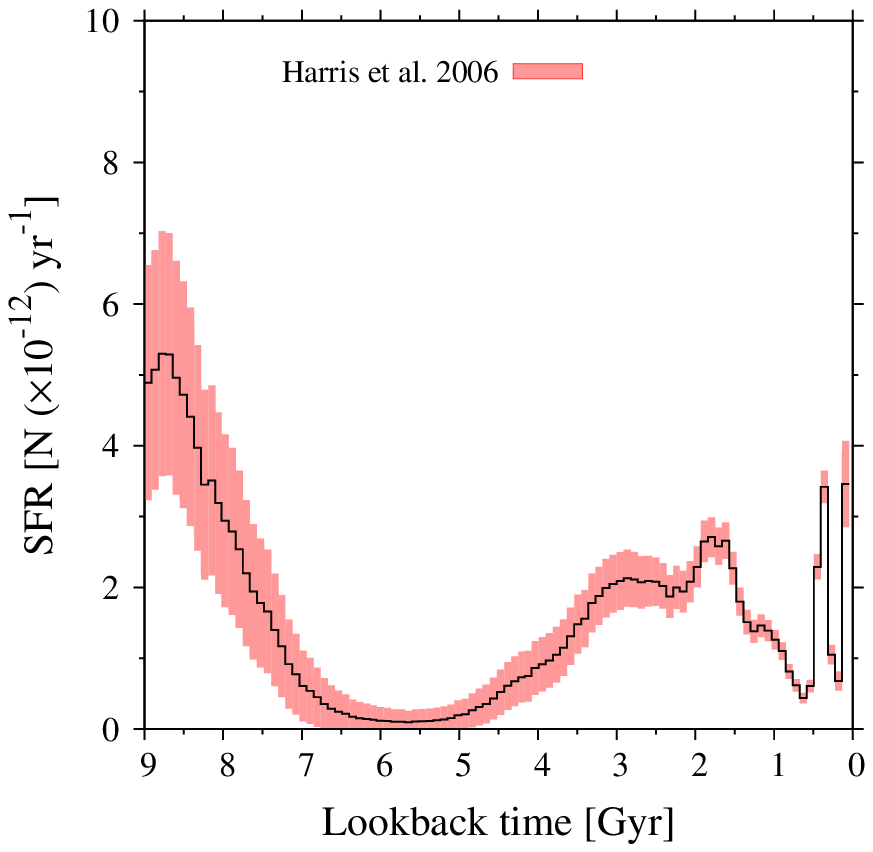}
\caption{Star formation rates recovered from two recent determinations of the Solar neighbourhood WDLF, that of RH11
and H06. The filled regions show the 1$\sigma$ uncertainty. In these tests, the algorithm converged in 11 and 28 steps.}
\label{fig:solneighbourhood}
\end{figure}

\begin{figure}[h!]
\plottwo{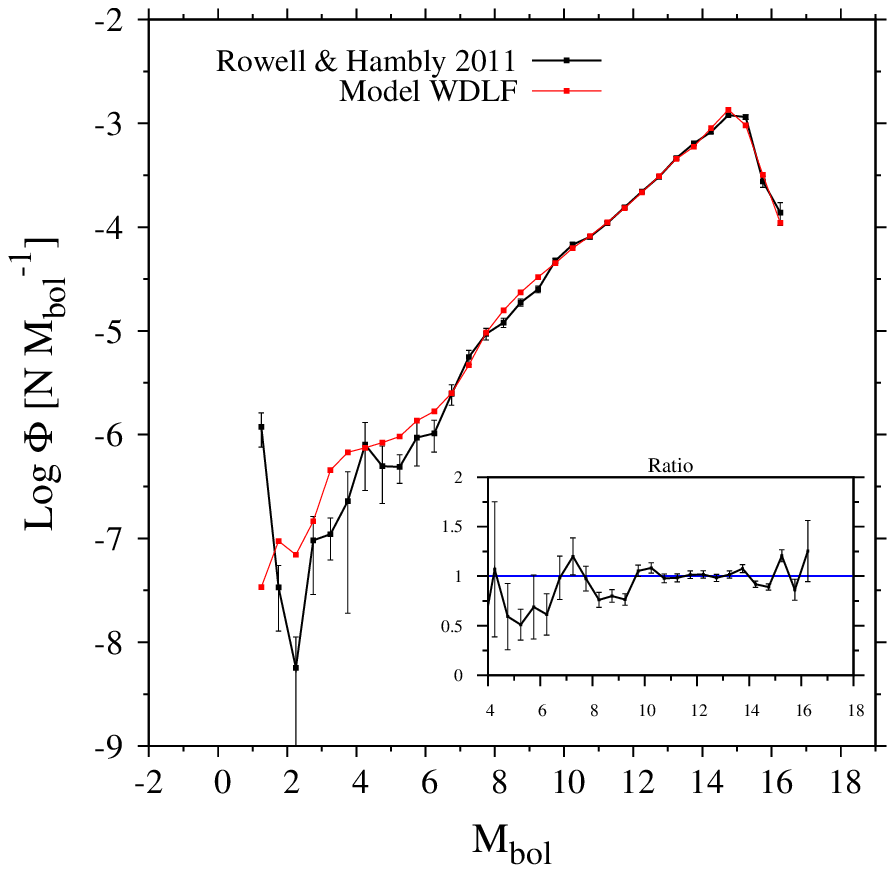}{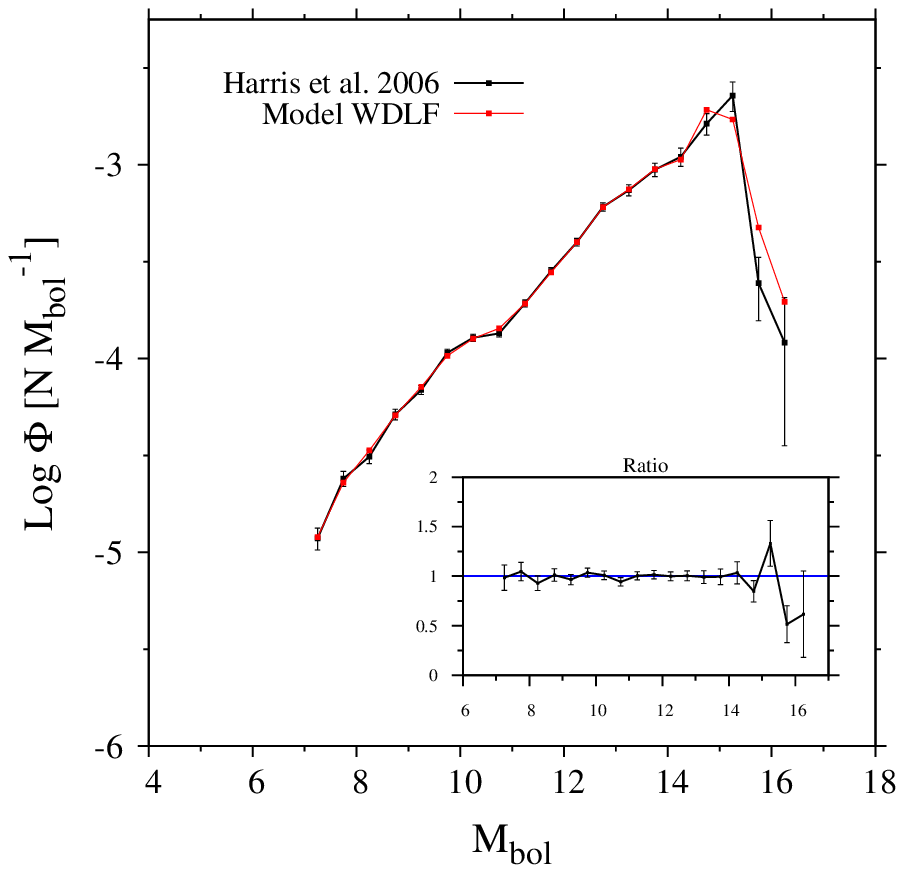}
\caption{Best fit WDLFs obtained from converged SFR models show a good fit to the observed WDLF in each case.}
\label{fig:obsWDLF}
\end{figure}

This algorithm has been applied to two recent measurements of the WDLF for the Solar neighbourhood: that of 
\citet{Rowell2011} and \citet{Harris2006} (hereafter RH11 and H06). The results are shown in figure \ref{fig:solneighbourhood}. Both 
SFRs show a similar form, being characterised by an early burst and a more recent peak at 2-3 Gyr in the past.
The difference in magnitude is due to the significant incompleteness ($\sim$50\%) of the RH11 sample with respect to that of H06.
In both cases, the maximum lookback time is fixed at 9 Gyr.
In figure \ref{fig:obsWDLF} the best fit synthetic WDLFs found on convergence of the algorithm are plotted over the observed WDLFs.
The inset panels show the ratio of the two functions. The WDLFs are fitted very well by the algorithm, particularly in the case of the H06 WDLF,
and there appears to be no significant over- or under-abundance of WDs that remains unaccounted for.
\section{Conclusion}
We have presented preliminary results of work to invert the WDLF. This represents a new method of analysing the star formation
history of the Solar neighbourhood and WD populations more generally, one that is essentially independent of existing inversion methods
that use main sequence stars, such as \citet{Hernandez2000} and \citet{Cignoni2006}. Application of the algorithm to the Solar
neighbourhood WDLF yields a SFR characterised by an early burst and a recent ($\sim2-3$ Gyr ago) peak.
Future development work will include making the maximum lookback time a free parameter. We also plan to compare results
for different sets of WD cooling models, which may turn out to be the largest uncertainty in the recovered SFR.
It would also be interesting to apply the method to other WD populations such as the thick disk, spheroid and clusters.
Single burst populations in particular would provide a useful benchmark.
\acknowledgements The author would like to thank Nigel Hambly, Gilles Fontaine and Enrique Garcia-Berro for useful discussions and feedback,
and the Space Technology Centre for financial support.
\bibliographystyle{asp2010}
\bibliography{rowell}

\begin{thebibliography}{}
\expandafter\ifx\csname natexlab\endcsname\relax\def\natexlab#1{#1}\fi
\expandafter\ifx\csname url\endcsname\relax
  \def\url#1{\texttt{#1}}\fi
\expandafter\ifx\csname urlprefix\endcsname\relax\def\urlprefix{URL }\fi
\providecommand{\eprint}[2][]{\url{#2}}

\bibitem[{Bedin et~al.(2010)Bedin, Salaris, King, Piotto, Anderson, \&
  Cassisi}]{Bedin2010}
Bedin, L., Salaris, M., King, I., Piotto, G., Anderson, J., \& Cassisi, S.
  2010, \apj, 708, L32

\bibitem[{{Bergeron} et~al.(2011){Bergeron}, {Wesemael}, {Dufour}, {Beauchamp},
  {Hunter}, {Saffer}, {Gianninas}, {Ruiz}, {Limoges}, {Dufour}, {Fontaine}, \&
  {Liebert}}]{Bergeron2011}
{Bergeron}, P., {Wesemael}, F., {Dufour}, P., {Beauchamp}, A., {Hunter}, C.,
  {Saffer}, R.~A., {Gianninas}, A., {Ruiz}, M.~T., {Limoges}, M.-M., {Dufour},
  P., {Fontaine}, G., \& {Liebert}, J. 2011, \apj, 737, 28

\bibitem[{Cignoni et~al.(2006)Cignoni, Degl'Innocenti, \& Moroni}]{Cignoni2006}
Cignoni, M., Degl'Innocenti, S., \& Moroni, P. 2006, \aap, 459, 783

\bibitem[{Dempster et~al.(1977)Dempster, Laird, \& Rubin}]{Dempster1977}
Dempster, A.~P., Laird, N.~M., \& Rubin, D.~B. 1977, Journal of the Royal
  Statistical Society. Series B (Methodological), 39, pp. 1

\bibitem[{Do \& Batzoglou(2008)}]{Do2008}
Do, C.~B., \& Batzoglou, S. 2008, Nature biotechnology, 26, 897

\bibitem[{Fontaine et~al.(2001)Fontaine, Brassard, \& Bergeron}]{Fontaine2001}
Fontaine, G., Brassard, P., \& Bergeron, P. 2001, \pasp, 113, 409

\bibitem[{Garc\'{\i}a-Berro et~al.(2010)Garc\'{\i}a-Berro, Torres, Althaus,
  Renedo, Lor\'{e}n-Aguilar, C\'{o}rsico, Rohrmann, Salaris, \&
  Isern}]{Garcia-Berro2010}
Garc\'{\i}a-Berro, E., Torres, S., Althaus, L.~G., Renedo, I.,
  Lor\'{e}n-Aguilar, P., C\'{o}rsico, A.~H., Rohrmann, R.~D., Salaris, M., \&
  Isern, J. 2010, \nat, 465, 194

\bibitem[{Harris et~al.(2006)Harris, Munn, Kilic, Liebert, Williams, von
  Hippel, Levine, Monet, Eisenstein, Kleinman, Metcalfe, Nitta, Winget,
  Brinkmann, Fukugita, Knapp, Lupton, Smith, \& Schneider}]{Harris2006}
Harris, H.~C., Munn, J.~A., Kilic, M., Liebert, J., Williams, K.~A., von
  Hippel, T., Levine, S.~E., Monet, D.~G., Eisenstein, D.~J., Kleinman, S.~J.,
  Metcalfe, T.~S., Nitta, A., Winget, D.~E., Brinkmann, J., Fukugita, M.,
  Knapp, G.~R., Lupton, R.~H., Smith, J.~A., \& Schneider, D.~P. 2006, \aj,
  131, 571

\bibitem[{Hernandez et~al.(2000)Hernandez, Valls‐Gabaud, \&
  Gilmore}]{Hernandez2000}
Hernandez, X., Valls‐Gabaud, D., \& Gilmore, G. 2000, \mnras, 316, 605

\bibitem[{Iben \& Laughlin(1989)}]{Iben1989}
Iben, I., \& Laughlin, G. 1989, \apj, 341, 312

\bibitem[{{Isern} et~al.(2008){Isern}, {Garc{\'{\i}}a-Berro}, {Torres}, \&
  {Catal{\'a}n}}]{Isern2008}
{Isern}, J., {Garc{\'{\i}}a-Berro}, E., {Torres}, S., \& {Catal{\'a}n}, S.
  2008, \apjl, 682, L109

\bibitem[{Knox et~al.(1999)Knox, Hawkins, \& Hambly}]{Knox1999}
Knox, R.~A., Hawkins, M. R.~S., \& Hambly, N.~C. 1999, \mnras, 306, 736

\bibitem[{Noh \& Scalo(1990)}]{Noh1990}
Noh, H.-R., \& Scalo, J. 1990, \apj, 352, 605

\bibitem[{{Oswalt} et~al.(1996){Oswalt}, {Smith}, {Wood}, \&
  {Hintzen}}]{Oswalt1996}
{Oswalt}, T.~D., {Smith}, J.~A., {Wood}, M.~A., \& {Hintzen}, P. 1996, \nat,
  382, 692

\bibitem[{{Rowell} \& {Hambly}(2011)}]{Rowell2011}
{Rowell}, N., \& {Hambly}, N.~C. 2011, \mnras, 417, 93

\bibitem[{{Tremblay} et~al.(2011){Tremblay}, {Bergeron}, \&
  {Gianninas}}]{Tremblay2011}
{Tremblay}, P.-E., {Bergeron}, P., \& {Gianninas}, A. 2011, \apj, 730, 128

\end{thebibliography}
\end{document}